\documentclass[aps,nofootinbib,notitlepage,longbibliography,twocolumn, superscriptaddress]{revtex4-1}
\usepackage{amsmath,amssymb}
\usepackage{slashed}
\usepackage{graphicx}
\usepackage{bm}
\usepackage{float}
\usepackage[T1]{fontenc}
\usepackage{multirow}
\usepackage[utf8]{inputenc}
\usepackage{gauss} 
\usepackage[normalem]{ulem}
\usepackage[top=1.0in,bottom=1.0in,left=1.0in,right=1.0in]{geometry}
\usepackage[colorlinks,linkcolor=blue,citecolor=blue]{hyperref}
\usepackage{subfig}
\usepackage{booktabs}
\usepackage{caption}
\newcommand{\be}{\begin{equation}}
\newcommand{\ee}{\end{equation}}
\newcommand{\bea}{\begin{eqnarray}}
\newcommand{\eea}{\end{eqnarray}}
\newcommand{\ba}{\begin{eqnarray}}
\newcommand{\ea}{\end{eqnarray}}

\newcommand{\ket}[1]{\left|#1\right\rangle}
\newcommand{\bra}[1]{\left\langle#1\right|}

\def\be{\begin{eqnarray}}
\def\ee{\end{eqnarray}}
\def\bea{\be}
\def\eea{\ee}

\def\roughly#1{\mathrel{\raise.3ex\hbox{$#1$\kern-.75em%
\lower1ex\hbox{$\sim$}}}}

\usepackage{lipsum}

\begin{document}

\title{Quasiparton distributions in massive QED2:\\
Toward quantum computation}
\author{Sebastian Grieninger}
\email[]{sebastian.grieninger@stonybrook.edu}
\affiliation{Center for Nuclear Theory, Department of Physics and Astronomy,
Stony Brook University, Stony Brook, New York 11794–3800, USA}
\affiliation{Co-design Center for Quantum Advantage (C2QA), Stony Brook University, Stony Brook, New York 11794–3800, USA}
\author{Kazuki Ikeda}
\email[]{kazuki.ikeda@stonybrook.edu}
\affiliation{Center for Nuclear Theory, Department of Physics and Astronomy,
Stony Brook University, Stony Brook, New York 11794–3800, USA}
\affiliation{Co-design Center for Quantum Advantage (C2QA), Stony Brook University, Stony Brook, New York 11794–3800, USA}

\author{Ismail Zahed}
\email[]{ismail.zahed@stonybrook.edu}
\affiliation{Center for Nuclear Theory, Department of Physics and Astronomy,
Stony Brook University, Stony Brook, New York 11794–3800, USA}

\date{\today}
\begin{abstract}
We analyze the quasiparton distributions of the lightest $\eta'$ meson 
in massive two-dimensional Quantum electrodynamics (QED2) by exact diagonalization. The Hamiltonian and boost operators are mapped onto spin qubits in a 
spatial lattice with open boundary conditions. The lowest excited state in the exact diagonalization is
shown to interpolate continuously between an anomalous $\eta'$ state at strong coupling,
and a non-anomalous heavy meson at weak coupling, with a cusp at the critical point. 
The boosted $\eta'$ state follows relativistic kinematics but with large deviations
in the luminal limit. The spatial quasiparton distribution function and amplitude  for the $\eta'$  state are computed numerically for increasing rapidity both at strong and weak coupling, and compared to the exact light front results. The numerical results from the boosted form of the spatial parton distributions, compare fairly with the inverse Fourier transformation
of the  luminal parton distributions, derived in the lowest Fock space approximation.
Our analysis points out some of the limitations facing the current lattice program
for the parton distributions.
\end{abstract}

\maketitle

\section{Introduction}
\label{SEC1}
Light cone distributions play a central role in the evaluation of hard inclusive and exclusive processes in Quantum Chromodynamics (QCD). Whenever factorization holds, a hard process is amenable to a perturbative contribution times pertinent parton distributions or fragmentation functions, e.g. Drell-Yan and jet
processes. Accurate light cone distribution functions are essential for the analyses of
many of the semi-inclusive processes being carried at the current JLAB facility, and planned at the future EIC~\cite{AbdulKhalek:2021gbh,Abir:2023fpo}. 

The parton distribution functions are valued on the light front and inherently non-perturbative, which make them inaccessible to standard Euclidean lattice formulations, except through few lowest moments. This major shortcoming has been circumvented by Ji's proposal~\cite{Ji:2013dva}, through the use of quasiparton distributions.  The leading twist light front partonic correlators are traded for equal-time correlators in 
increasingly boosted hadronic states, which are then matched perturbatively
on the light cone~\cite{Zhang:2017bzy,Ji:2015qla,Bali:2018spj,Alexandrou:2018eet,Izubuchi:2019lyk,Izubuchi:2018srq}.

The quasiparton distribution matrix elements calculated in a fixed size Euclidean lattice QCD have been shown to  match those obtained through LSZ reduction in continuum Minkowski QCD to all orders in perturbation theory~\cite{Briceno:2017cpo}. Some  variants of this formulation have also been proposed as pseudo distributions~\cite{Radyushkin:2017gjd} and lattice cross sections~\cite{Ma:2014jla}. Recently, an increasing number of QCD lattice collaborations have implemented some of these ideas, and succeeded in numerically extracting the light cone parton distributions,
modulo the shortcomings inherent to a finite lattice.

To understand some of these shortcomings, quasiparton distribution functions have been analyzed in two-dimensional QCD in leading and next-to-leading contributions in the large $N_c$ limit~\cite{Ji:2018waw}. 
The results confirm the soundness of the approach, away from the critical 
regions in parton-x $x=0,1$, where the method breaks down. In this work, we would like to revisit the analysis of the quasiparton distributions 
in two-dimensional massive QED utilizing a spin formulation of the lattice Schwinger model and solving it by exact diagonalization as an exploratory study for a future quantum computation. The
purpose is to understand the feasibility and numerical issues that follow
from this digitization. We note that a similar approach, has been recently carried
for the parton distributions for a schematic Nambu-Jona-Lasinio (NJL) model in two-dimensions~\cite{Li:2021kcs,Li:2022lyt}. In the context of quantum computations, partonic observables were previously considered in~\cite{Lamm:2019uyc,Perez-Salinas:2020nem,Briceno:2020rar}. Finally, simulating quantum field theories in the light-front formulation on quantum computers was explored in~\cite{Kreshchuk:2020dla,Echevarria:2020wct}.

In this work massive QED2 is formulated on a spatial lattice with continuous time, using Kogut-Suskind fermions~\cite{Kogut:1974ag,Susskind:1976jm}. The fermion fields are eliminated with the help of a Jordan-Wigner transformation~\cite{Jordan:1928wi}. While these procedures are by now
standard, their application to understand the partonic content of physical states in a gauge
theory with massive QED2 as a protype, is new. More specifically,
our paper contains a number of new results: 1)  We show anew the central role played by the boost operator in the analysis of the light front wavefunctions in real time; 2) We show anew how the low-lying physical spectrum of a gauge theory transforms under increasing boosts, with a quantitative  description of the numerical limitations in the luminal limit in QED2; 3) We provide a new Hamiltonian formulation to the concept of quasi-PDFs in real-time, in a gauge theory with QED2 as an example; 4) We use the qubit form of the QED2
Hamiltonian and boost operators to extract anew the spatial quasidistributions in QED2 in real time, and study their dependence on the boost transformation with a comparison to the
expected exact results. These are essential new theoretical steps and checks,  to enforce in any attempt on the road towards quantum computation of partonic distributions in hadronic physics.

Quantum  computing and quantum simulation offer unique capabilities that vastly surpass what classical computation alone can achieve, particularly in addressing specific challenges in  Nuclear Physics~\cite{Beck:2023xhh}.
A key challenge in quantum simulation/computation of gauge theories, is a scalability of the system size. The  ground state of general one-dimensional spin chains can be solved using the  Matrix Product State (MPS) approximation~\cite{1992CMaPh.144..443F,PhysRevB.55.2164,PhysRevLett.75.3537}, which underlines the Density Matrix Renormalization Group (DMRG)~\cite{PhysRevLett.69.2863,RevModPhys.77.259} algorithm. A number of challenging properties of the ground state of the Schwinger model have been studied with quantum computers~\cite{Farrell:2023fgd,Farrell:2024fit} and simulators~\cite{Florio:2024aix} for a large number of qubits (up to hundreds qubits). However, the  non-trivial dynamics of the excited states has received less attention, although some basic properties of the first excited state were studied~\cite{PhysRevResearch.4.043133,Banuls:2013jaa}. To open a new directional study in massive QED2, as a testing ground for four dimensional
QCD, we will focus on the boosted properties of the first excited state
and its partonic content. Our results will provide interesting and important benchmarks towards a quantum computation in QCD. Moreover our setup can be extended to a large system size by using more advanced MPS methods, to approximate the excited states~ \cite{Banuls:2013jaa,PhysRevLett.113.091601}.

The outline of the paper is as follows: in section~\ref{SEC2}, we briefly review the
general features of the strong and weak coupling regime of massive QED2. In section~\ref{SEC3}, we recall the light front equation for the lowest meson state
in the lowest Fock state approximation, and its solutions.
In section~\ref{SEC5}, we introduce the boost operator, and  
map massive QED2 including the boost operator onto a spin system using standard techniques. A detailed analysis of the deviation from the luminal limit for the boosted state is given. We also show how the boost operator can be traded for a "time" evolved
spatial correlator directly on the light front in the luminal case.
In section~\ref{SEC6}, we detail the numerical results for the parton distribution functions and amplitudes obtained from the exact diagonalization, using the boosted as well as the "time" evolved forms, both for strong and weak coupling. The results are compared   to the exact
continuum result. Our conclusions are in section~\ref{SEC7}.

\section{Massive QED2}
\label{SEC2} 
The two-dimensional Quantum Electrodynamics (QED2), also known as the Schwinger model~\cite{Schwinger:1962tp}, exhibits a variety of non-perturbative phenomena familiar from four-dimensional 
gauge theories. The extensive interest in QED2 stems from the fact that it bears much in common with two-dimensional QCD since the Coulomb law is confining in two dimensions. As a result, the QED2 spectrum involves only chargeless excitations. Remarkably, the vacuum state is characterized by a non-trivial chiral condensate and topologically active tunneling configurations.

QED2 with massive fermions is described by~\cite{Schwinger:1962tp,Coleman:1976uz}
\bea
\label{A1}
S=\int d^2x\,\bigg(\frac 14F^2_{\mu\nu} +\frac{\theta \tilde F}{2\pi}+\overline \psi (i\slashed{D}-m)\psi\bigg)
\eea
with $\slashed{D}=\slashed{\partial}-ig\slashed{A}$. The bare fermion mass is $m$ and the coupling $g$ has mass dimension. At weak coupling with $m/g>1/\pi$, the spectrum is composed of heavy mesons, with doubly degenerate C-even and C-odd vacuua at $\theta=\pi$.  At strong coupling  with $m/g<1/\pi$, the spectrum is  composed of light mesons and baryons, with a C-even vacuum whatever $\theta$. More specifically:
\\
\\
{\bf Strong coupling $\frac mg\ll \frac 1\pi$:}
\\
The squared rest mass is
\bea
m_\eta^2=m_S^2+m_\pi^2=\frac {g^2}\pi-\frac{m\langle \overline{\psi}\psi\rangle_0}{f^2},
\eea 
with $f=1/\sqrt{4\pi}$ the $\eta'$ decay constant~\cite{Grieninger:2023ufa}. 
The vacuum chiral condensate is finite
in the chiral limit with
$\langle\overline\psi \psi\rangle_0=-\frac{e^{\gamma_E}}{2\pi}m_s$
~\cite{Sachs:1991en,Steele:1994gf}, 
where $\gamma_E=0.577$ is Euler constant. Hence
\bea
\label{WC1}
\frac{m_\eta}{m_s}=
\bigg(1+2e^{\gamma_E}\frac{m}{m_s}\bigg)^{\frac 12}\approx 
1+e^{\gamma_E}\frac m{m_s}
\eea
in the strong coupling regime.
\\
\\
{\bf Weak coupling $\frac mg\gg \frac 1\pi$:}
\\
The $\eta$-mass is expected to asymptote
$m_\eta\rightarrow 2m$. Note that at weak coupling (\ref{WC1}) we have $e^{\gamma_E}\approx 1.78$ which is close to 2,
hence allowing a smooth and continuous transition in the eta mass from weak to strong coupling.

\section{Light front wavefunctions}
\label{SEC3}
Modulo the U(1) anomaly, the light front Hamiltonian associated with QED2 is very 
similar to QCD2. Massless QED2 is exactly solvable with a single bosonic excitation of mass $m_s$ in the spectrum.  On the contrary, massive QED2 is not exactly solvable, and is characterized by multi-boson states. 

In the 2-particle Fock approximation, massless QED2 admits a ground state and a tower of spurious states that disappear in the continuum when higher
Fock states contributions are included~\cite{Bergknoff:1976xr}. Massive QED2
in the 2-particle Fock space approximation admits
a ground state and a tower of multi-meson states that are not spurious. The ground state is largely a 2-particle Fock contribution while the first excited state already contains a large 4-particle Fock space contribution (meson-meson bound state)~\cite{Mo:1992sv}.


With this in mind, the light front wavefunctions $\varphi_n(\zeta)$ in the 2-particle Fock-space
approximation follow from a similar Bethe-Salpeter derivation. More specifically, the
light front wavefunctions solve~\cite{Bergknoff:1976xr}
\begin{widetext}
\bea
\label{TH1}
M_n^2\varphi_n(\zeta)=\frac 12 m_s^2\int_{-1}^1\,d\zeta'\,\varphi_n(\zeta')+\frac {4m^2}{1-\zeta^2}\varphi_n(\zeta)-2m_s^2\,{\rm PP}\int_{-1}^1\,d\zeta'\,\frac{\varphi_n(\zeta')-\varphi_n(\zeta)}{(\zeta'-\zeta)^2}
\eea
\end{widetext}
with $\zeta P$  the symmetric  momentum fraction of the partons. Here PP is short for the principal part. Eq. (\ref{TH1}) is the 't Hooft equation~\cite{tHooft:1974pnl}, modulo the U(1) anomaly contribution (first term on the right hand side). 
The longitudinal kinetic contribution (second term on the right hand side)  is singular at $\zeta=\pm 1$,
forcing the light front wavefunction to vanish 
at the edges $\varphi_n(\pm 1)=0$.
Note that the 2-particle Fock-contribution is the leading contribution in the limit of a large number of colors in QCD2.

In the massless limit with $m\rightarrow 0$, the 
spectrum is that of a single massive boson
\bea
\label{PHIM0}
\varphi_\eta(\zeta )=\varphi_0(\zeta)\rightarrow \frac 1{\sqrt 2}\theta(1-\zeta^2),\qquad M_0^2\rightarrow m_s^2.
\eea
 For small $m$,  the lowest solution vanishes at the end-points as powers of $m$, 
\bea
\label{SOL0}
\varphi_\eta(x)=\frac{C_\eta}{f} (1-\zeta^2)^\beta
\eea
with $\beta$ fixed by~\cite{tHooft:1974pnl}
\bea
\frac{m^2}{m_s^2}=1-\pi\beta{\rm cotan}(\pi\beta).
\eea
In the massive limit with $m\gg m_s$, the solution is peaked around $x=\frac 12$, with $M\approx 2m$.

In general, (\ref{TH1}) admits a tower of excited states for finite $m/m_s$ since the PP part in (\ref{TH1}) is strictly confining. They are given semi-classically by
\bea
\label{SEMI}
\varphi_n(x)=\sqrt{2}\,{\rm sin}\bigg((n+1)(1+\zeta)\frac \pi 2\bigg).
\eea
Because of screening, which is not suppressed in massive QED2, the excited
mesons states decay whenever $M_n>2M_0$. 
The lowest state parton distribution function is
of the form
\bea
\label{QM0}
q_\eta(\zeta )={\rm sgn} (\zeta)|\varphi_\eta (|\zeta| )|^2.
\eea
with $\rm sgn (\zeta)$  the signum function. Since $\eta'$ is flavor neutral, the 
fermion and anti-fermion distribution functions are identical.

The integral equation (\ref{TH1}) is a relative of the 't Hooft equation and can be solved by variational or other numerical methods. Here, we use a direct discretization into an matrix eigenvalue equation as for the 't Hooft equation~\cite{Zubov:2016bqs}. For that we define the rescaled masses 
$$\mu^2=M^2/m_s^2\qquad\alpha=m^2/m_s^2$$ 
so that (\ref{TH1}) reads
\begin{widetext}
\bea
\label{TH1X}
\mu^2\varphi(\zeta)=\frac 12
\int_{-1}^1d\zeta'\,\varphi(\zeta')+
\frac{4\alpha}{1-\zeta^2}\varphi(\zeta) 
-2{\rm PP}\int_{-1}^1\,d\zeta'\,\frac{\varphi(\zeta')-\varphi(\zeta)}{(\zeta-\zeta')^2}.
\eea
\end{widetext}
subject to the end-point conditions $\varphi(\pm 1)=0$. 
The solutions follow numerically by expanding in a complete basis, the details of which are summarized in Appendix~\ref{APPNUM}. A more standard but straightforward x-representation is 
summarized in Appendix~\ref{APPX}.

In Fig.~\ref{qfunct} we show typical solutions in the strong and weak coupling regimes.
We will refer to these solutions  as ``exact solution'' even though the solutions are approximated, in the sense that we cut the expansion in the Jacobi basis at a finite number of Jacobi polynomials. However, higher Jacobi polynomials are rapidly suppressed.

\begin{figure}
    \centering
    \includegraphics[width=0.99\linewidth]{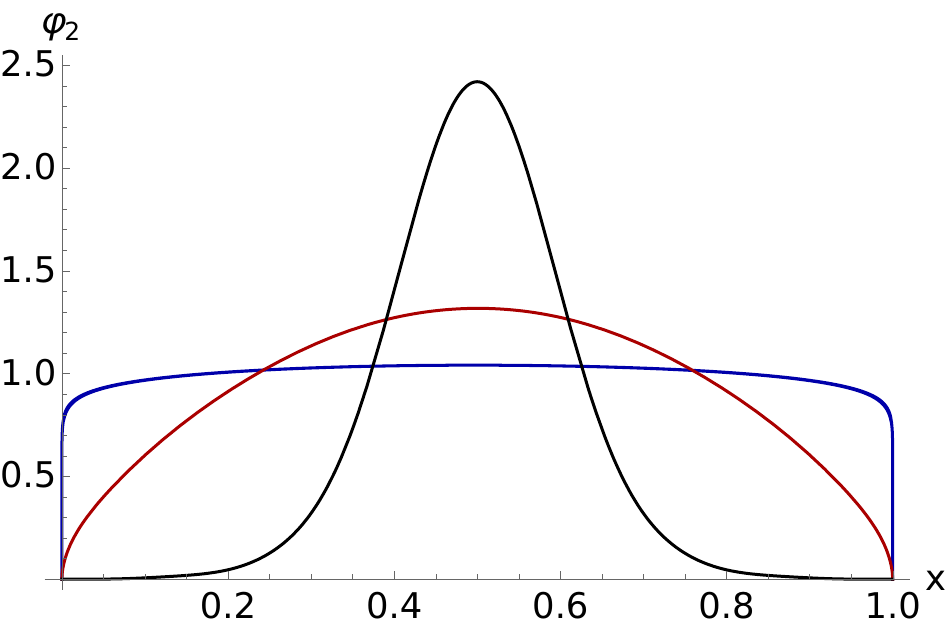}
    \caption{Solution to \eqref{TH1X} for $\beta=0.1\sqrt{3}/\pi  $ (blue), $\beta=\sqrt{3}/\pi$ (red), $\beta=10\sqrt{3}/\pi  $ (black) using 13 Jacobi polynomials.}
    \label{qfunct}
\end{figure}

\section{Boosted quasipartonic functions}
\label{SEC5}
To quantify the partonic distributions in the lowest meson $\eta'$ in massive QED2,
we will make use of the appropriate Ji's quasi-distributions~\cite{Ji:2013dva}, in a boosted state.
This construction allows for testing the numerical limitations when addressing the 
ultra-relativistic and large spatial volume limits. We will present two equivalent forms of the quasi-distribution functions, one using explicitly the boost operator and the other trading it for the "time" evolved operator.

\subsection{Boost operator in QED2}
\label{SEC4}
For $\theta=0$ and in the temporal gauge $A^0=0$, the Hamiltonian and boost operators associated to (\ref{A1}) read
\bea
\label{HBX}
\mathbb H&=&\int dx\,
\bigg(\frac 12 E^2+\psi^\dagger (i\alpha D_x+m\gamma^0)\psi\bigg),\nonumber\\
\mathbb K&=&\int dx\,x\,
\bigg(\frac 12 E^2+\psi^\dagger (i\alpha D_x+m\gamma^0)\psi\bigg).
\eea
Note that both operators develop CP odd contributions under the substitution $m\rightarrow me^{i\gamma^5\theta}$ for a finite vacuum angle. The operators (\ref{HBX}) satisfy
the Poincare algebra
\bea
\label{KHP}
[\mathbb K, \mathbb H]&=&i\mathbb P,\\
\left[\mathbb K, \mathbb P\right]&=&i\mathbb H,
\eea
with the momentum operator $\mathbb P$.

Physically boosted states are defined as
\bea
\label{BOOSTSTATE}
|\eta(\chi)\rangle =e^{i\chi \mathbb K}|\eta(0)\rangle,
\eea
with $\chi$ the rapidity in the x-direction
\bea
\chi=\frac 12 {\rm ln}\bigg(\frac{1+v}{1-v}\bigg).
\eea
If $\eta'$ is the lowest massive meson in the spectrum 
at strong coupling, with rest mass (normal ordering assumed)
\bea
:\mathbb H:|\eta(0)\rangle = m_\eta |\eta(0)\rangle,
\eea
then the boosted state is relativistic
\bea
:\mathbb H:|\eta(\chi)\rangle &=& m_\eta\, {\rm cosh}\chi|\eta(\chi)\rangle,\nonumber\\
:\mathbb P:|\eta(\chi)\rangle &=& m_\eta\, {\rm sinh}\chi|\eta(\chi)\rangle,
\eea
with 4-momentum $p^\mu=\gamma m_\eta (1, v)$,  and $\gamma=\, {\rm cosh}\chi=1/\sqrt{1-v^2}$. Note that  on the light cone
$$P^\pm =\frac 1{\sqrt 2} \gamma m_\eta (1\pm v)=\frac 1{\sqrt 2} m_\eta\,e^{\pm \chi}.$$

\subsection{Boosted quasi-distributions}
With this in mind, we 
define the distribution amplitude for the boosted $\eta'$ meson as
\begin{widetext}
\bea
\label{QPDA}
\varphi_\eta(x, v)=
\frac {1}{2if}
\int_{-\infty}^{+\infty} \frac{dz}{2\pi}\,
e^{-iz\zeta p^1}
\langle 0|\overline\psi(0,- z)[- z, + z]_S\gamma^+\gamma^5\psi(0,+ z)\,e^{i\chi \mathbb K}\,|\eta(0 )\rangle,
\eea
with $p^1=\gamma m_\eta v$  and  
$\zeta=2x-1$ with $x$ the parton fraction.
The link variable $[-z,+z]_S$ is along the spatial direction, 
and $f=1/\sqrt{4\pi}$ the 2-dimensional eta decay constant. 
Here $\gamma^+=\gamma^0+\gamma^1$, $\gamma^5=\gamma^0\gamma^1$,  
with $\gamma^0=\sigma^3$ and $\gamma^1=i\sigma^2$. Note that in 2-dimensions
$\overline\psi \gamma^+\psi=\overline\psi\gamma^+\gamma^5\psi$. Similarly, 
the partonic distribution function is defined as
\bea
\label{QPDF}
q_\eta(x, v)=\frac 12
\int_{-\infty}^{+\infty} \frac{dz}{2\pi}\,
e^{- iz\zeta p^1}
\langle \eta(0)|\,e^{- i\chi \mathbb K}\,\overline\psi(0,- z)[- z,+ z]_S\gamma^+\gamma^5\psi(0,+ z)\,e^{i\chi \mathbb K}\,|\eta(0 )\rangle.
\eea
\end{widetext}
Here (\ref{QPDA}-\ref{QPDF}) are partonic functions defined at equal time for a fixed boost, which reduce to the
light front partonic functions in the large rapidity limit $\chi\gg 1$. 
They  are Ji's quasipartonic functions, and therefore have support on $[0,1]$ only in the large rapidity limit. 
Note that both (\ref{QPDA}-\ref{QPDF}) normalize to 1 in this limit.

\subsection{"Time" evolved quasi-distributions}
The boost operator in the quasi-distribution, can be traded for a Hamiltonian evolution.
In the luminal case, it reduces to a direct evaluation of the parton distribution functions on the light front with Minkowski signature. This is possible since QED2 is UV finite. To show this, 
we use the  boost identities
\bea
\label{LOR}
e^{-i\chi \mathbb K}\psi(0,+ z)e^{i\chi \mathbb K}&=&S[v]\,
\psi(- \gamma v z, + \gamma z),\nonumber\\
e^{-i\chi \mathbb K}\overline\psi(0,- z)e^{i\chi \mathbb K}&=&
\overline\psi(+ \gamma v z, -\gamma z)\,S[v]^{-1},
\eea
with $S[v]=e^{\frac 12\chi\gamma^5}$. Inserting (\ref{LOR}) into (\ref{QPDA}) yields
\begin{widetext}
\bea
\label{QPDA2}
\varphi_\eta(x, v)=
\frac {1}{2if}
\int_{-\infty}^{+\infty} \frac{dz}{2\pi}\,
e^{- iz\zeta m_\eta v}
\langle 0|\overline\psi(+ vz,- z)[- z,+ z]_L\Gamma_v\psi(- vz,+ z)|\eta(0)\rangle.
\eea
\end{widetext}
after a shift $\gamma z\rightarrow z$ with $\Gamma_v=-(\gamma^-+ v\gamma^+)$ and 
the gauge link $[-z,+z]_L$ is now along the light cone. Note that the standard spin source $\Gamma_v\rightarrow \gamma^+$ in (\ref{QPDA2}) is obtained by using   $\gamma^+\gamma^5\rightarrow \gamma^0$ in (\ref{QPDA}).
We have traded the boosted $\eta'$ state for 
a fermionic bi-local correlator with light-cone support for $v\rightarrow 1$. In this limit, 
Ji's quasi-distribution is the light-front distribution
for the $\eta'$ meson,  expressed in the $\eta'$ rest frame, for both choices of $\Gamma_v$ quoted. No matching is required since massive QED2 is finite in the UV modulo normal ordering.
Also note that the boost transformation  turns the spatial gauge link light-like at the end-points.

Using the "time" identities
\bea
\psi(- v z, +  z)&=&
e^{- ivz\mathbb H}\psi(0,+ z)e^{+ ivz \mathbb H},
\nonumber\\
\overline\psi(+ v z, -  z)&=&
e^{+ ivz\mathbb H}\psi(0,- z)e^{- ivz \mathbb H},
\eea
we can recast (\ref{QPDA2}) as
\begin{widetext}
\bea
\label{QPDA3}
\varphi_\eta(x, v)=
\frac {1}{2if}
\int_{-\infty}^{+\infty} \frac{dz}{2\pi}\,
e^{- iz(\zeta-1) m_\eta v}
\langle 0|\overline\psi(0,- z)[- z,+ z]_L\Gamma_v\,e^{- i2vz{\mathbb H}}\,\psi(0,+z)|\eta(0)\rangle.
\eea
The boost operator in (\ref{QPDA2}) has been traded for the real 'time evolution operator', albeit with  the effective "time" of $t_z=2vz$. 
A rerun of the same arguments for the parton distribution function (\ref{QPDF}), yields
\bea
\label{QPDF3}
q_\eta(x, v)=
\frac {1}{2}
\int_{-\infty}^{+\infty} \frac{dz}{2\pi}\,
e^{- iz\zeta m_\eta v}
\langle \eta(0)|\overline\psi(0,- z)[- z,+ z]_L\Gamma_v\,e^{- i2vz{\mathbb H}}\,\psi(0,+ z)|\eta(0)\rangle,
\eea
\end{widetext}
with  the normalization 
$$\int_0^1 dx\,p_\eta(x,v)=1.$$
This "time" evolved form for $v=1$ was noted in~\cite{Li:2021kcs,Li:2022lyt}.
The advantage of (\ref{QPDA3}-\ref{QPDF3}) over  (\ref{QPDA2}),
is their manifest finiteness as $v\rightarrow 1$.
Below, we will use the digitalization of both (\ref{QPDA2}) and (\ref{QPDA3}-\ref{QPDF3}) for a cross check on their accuracy on a finite grid of qubits.

\section{Latticized Kogut-Susskind
Hamiltonian}
\label{SEC6}
We will map our Hamiltonian onto a lattice spin system of length $L=Na$ to make it suitable for exact diagonalization.
Using staggered fermions, the map for the two-dimensional massive
Schwinger model and gauge fixing using Gauss law is standard~\cite{Banks:1975gq}. Here, we only detail
the mapping of the boost operator for completeness. Throughout the work, we obtained the excited states by exact-diagonalization of the Hamiltonian~\eqref{eq:Ham}. One can implement such excited states on a quantum computer by 
using a Variational Quantum Deflation (VQD) algorithm~\cite{Higgott:2018doo}. 

\subsection{Digitized boost operator}
More specifically, the gauge field contribution to the boost operator is
\bea
\mathbb K_G=\frac 12 g^2a\sum_{n=0}^{N-1} nL_n^2,
\eea
where $L_n$ are link operators~\cite{Banks:1975gq}. Each staggered Dirac fermion at site $n$ 
is assigned a bi-spinor  with upper
component on even sites, and lower component on  odd sites
\bea
\psi(0,z=na)=\frac 1{\sqrt a}
\begin{pmatrix}
   \psi_e(n) \\
   \psi_o(n)
\end{pmatrix}
=\frac 1{\sqrt a}
\begin{pmatrix}
   \varphi_{n: {\rm even}} \\
   \varphi_{n+1: {\rm odd}}
\end{pmatrix},\nonumber\\
\eea
with $0\leq n\leq N-1$. The massive staggered fermion contribution to the boost operator is 
\bea
\mathbb K_F=&&\frac 1{2}\sum_{n=0}^{N-1} in(\varphi^\dagger_n\varphi_{n+1}-\varphi^\dagger_{n+1}\varphi_n)\nonumber\\
&&+ma \sum_{n=0}^{N-1}\,(-1)^n n
\varphi^\dagger_n\varphi_n.
\eea
Fermion bilocals with the gauge link set to 1
(in 2-dimensions the gauge field can be eliminated by fixing
the gauge and resolving Gauss law), are digitized as
\bea
&&\overline\psi( {0,-z})(\gamma^0+\gamma^1)\gamma^5\psi(0,+z)=\nonumber\\
&&\frac 1a (\psi_e^\dagger(-n) +\psi_o^\dagger(-n))
(\psi_e(+n) +\psi_o(+n)).
\eea

\begin{figure}
    \centering
    \includegraphics[width=1\linewidth]{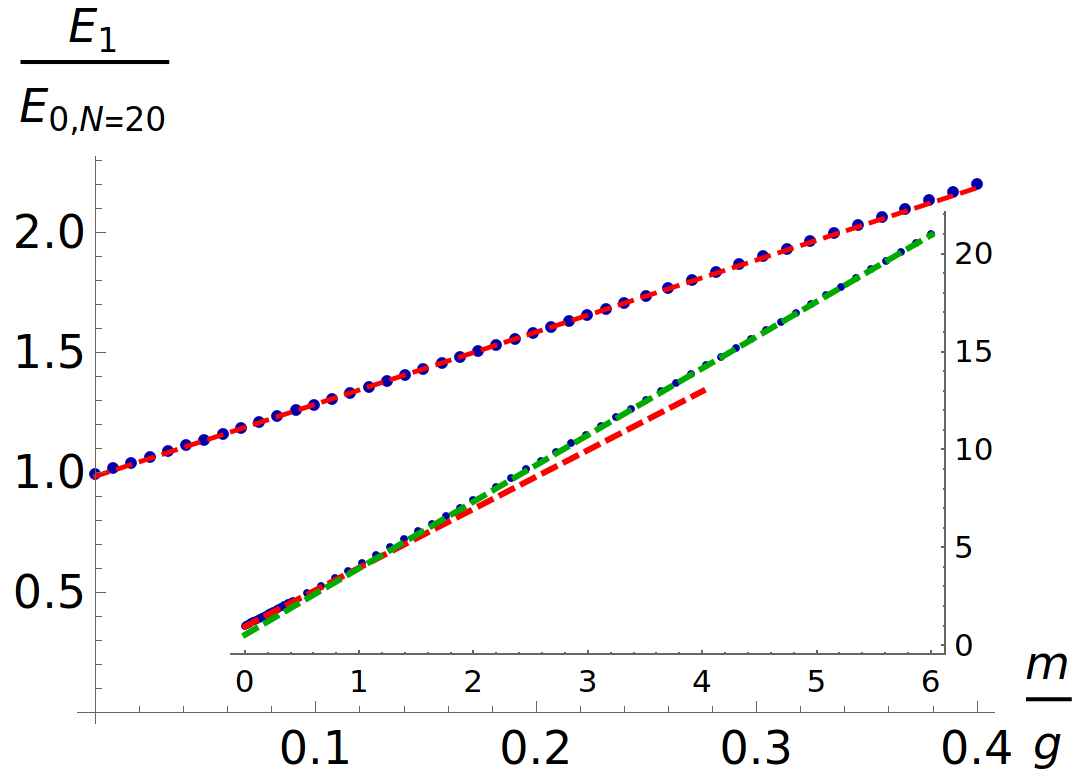}
    \caption{
    Normalized mass gap as a function of the improved mass $m_\text{lat}$ with $g=a=1$ and $N=20$. The red-dashed line is a fit to (\ref{FIT1}) and the green-dashed line in the inset is a fit to (\ref{FIT2}).}
    \label{fig:SPEC1}
    \end{figure}
\begin{figure}
    \includegraphics[width=1\linewidth]{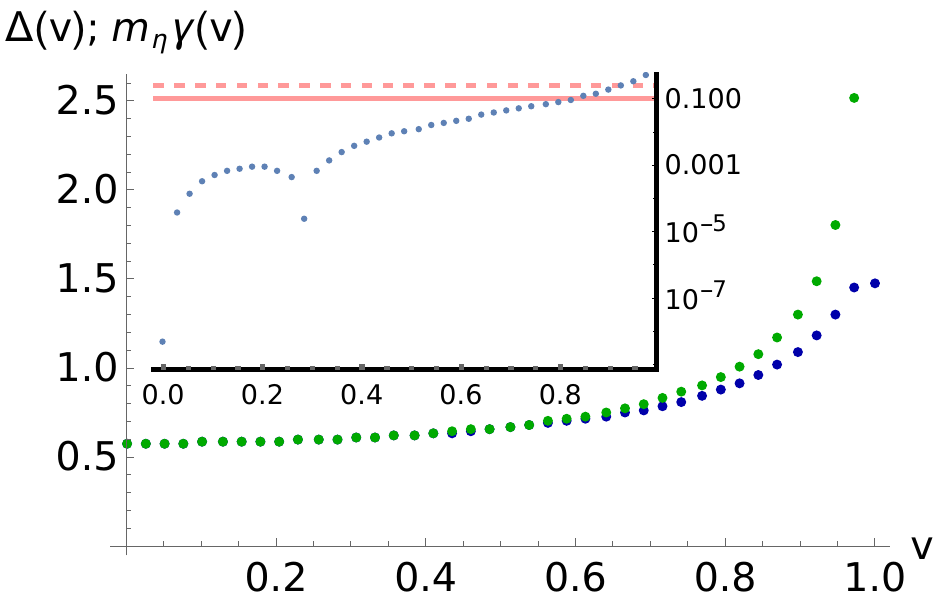}
    \caption{ Relativistic dispersion of the $\eta'$ state in dotted-blue (exact diagonalization)
    and in dotted-green (exact). We fixed $N=24$ and the improved mass as $m_{\text{lat}}=-0.125 g^2a$ ($a=1=g)$. The symmetric relative error is shown in the inset
    in dotted-blue. The red line indicates an error in excess of $10\%$ (at around $v\gtrsim0.83$), while the dashed-red line implies an error in excess of  $20\%$ (at around $v\gtrsim0.91$).}
    \label{fig:boosted}
\end{figure}

\begin{figure}
    \centering
    \includegraphics[width=1\linewidth]{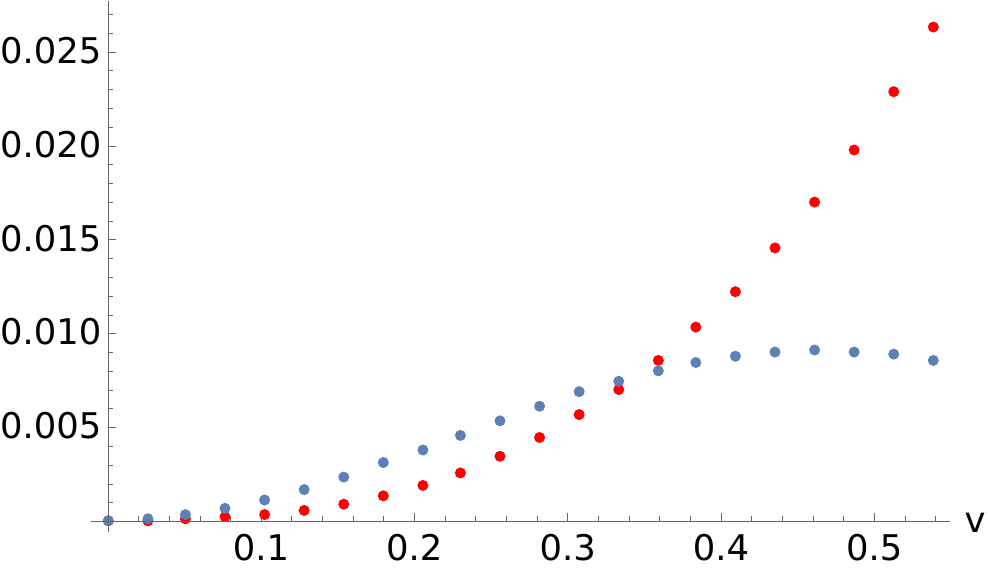}
    \caption{Symmetric relative error of the boost (red dots) and $\langle\eta(0)|0(v)\rangle$ (blue dots) for $N=24$ and $m_\text{lat}=0.1/\sqrt{\pi}-0.125 g^2a (a=g=1)$.}
    \label{fig:boostederr}
\end{figure}

The chiral fermions map onto $\psi_e(n)=\varphi_n$  (even sites) and
$\psi_o(n)=\varphi_n$ (odd sites) with~\cite{Kogut:1974ag,Susskind:1976jm,Jordan:1928wi,Banks:1975gq}
\bea
\varphi_n&=&\prod_{m<n}[+iZ_m]\frac 12 (X_n-iY_n),\nonumber\\
\varphi^\dagger_n&=&
\prod_{m<n}[-iZ_m]\frac 12 (X_n+iY_n).
\eea
The full boost operator maps onto
\bea
\mathbb K\rightarrow 
&&\frac 12 g^2a\sum_{n=0}^{N-1} nL_n^2\nonumber\\
&&+\frac 1{4}\sum_{n=0}^{N-1}\,n
\left(X_nX_{n+1}+Y_nY_{n+1}\right)\nonumber\\
&&+\frac{ma}2
\sum_{n=0}^{N-1}\,(-1)^n n\,(1+Z_n).
 \eea
and the Hamiltonian maps onto \cite{Kharzeev:2020kgc,deJong:2021wsd,PhysRevD.103.L071502,Florio:2023dke,PhysRevD.108.L091501,Barata:2023jgd}
\begin{align}
\begin{aligned}
\label{eq:Ham}
\mathbb{H}\to&
     \frac{1}{4a}\sum_{n=1}^{N-1}\left(X_n X_{n+1}+Y_n Y_{n+1}\right)
\\&
     +\frac{m}{2}\sum_{n=1}^N(-1)^n Z_n+\frac{a g^2}{2}\sum_{n=1}^{N-1}L^2_n,
 \end{aligned}
 \end{align}
where $L_n$ is the local electric field operator 
\begin{equation}
    L_n=\sum_{j=1}^n\frac{Z_j+(-1)^j}{2}.
\end{equation}

    \begin{figure}
     \includegraphics[width=1\linewidth]{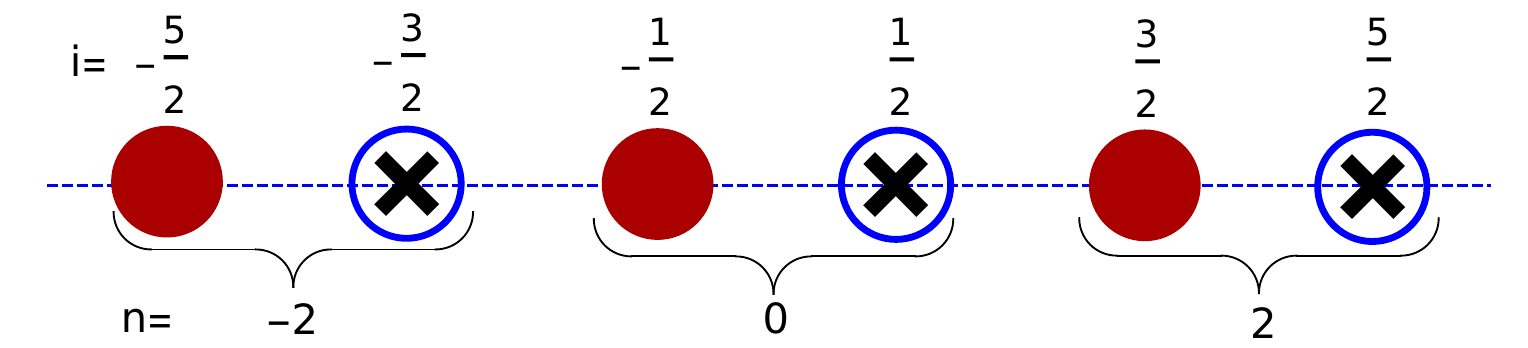} \includegraphics[width=0.6\linewidth]{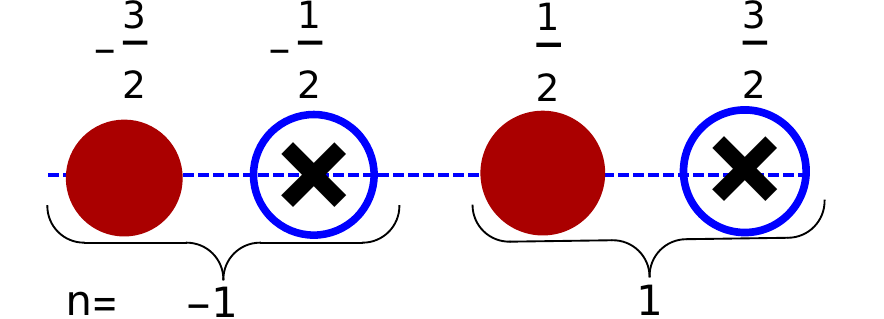}
    \caption{Configuration of the staggered fermions for $N/2$ odd (upper plot for $N=6$) and $N/2$ even (lower plot for $N=4$). To compute the correlation functions such as \eqref{eq:QPDF}, we symmetrically pair the first underbracket with the last, the second with the second to last etc.}
    \label{fig:lattice}
\end{figure}

\subsection{Digitized boosted $\eta'$-state}
The discrete lattice version of the boost operator can be tested on the $\eta'$ state at rest to quantify
the numerical deviations  from the relativistic result in the continuum. For that, we start by setting the scale $g=1$, and recording all other scales in units of $g$.
For now, we fix the number of qubits $N=20$ with a spatial lattice spacing $a=1$. For better numerical accuracy in the Schwinger model, we follow~\cite{Dempsey:2022nys} and evaluate the expressions as a function of the improved mass
\begin{equation}\label{eq:mlat}
    m_\text{lat}=m-\frac{1}{8}g^2a\rightarrow m-\frac 18.
\end{equation}
The mass gap in a finite spatial box of length $L=Na\rightarrow N$, receives
finite size corrections $E_0=\sqrt{m_s^2+\pi^2/L^2}$ with $L=N\cdot a$, with $m_s^2=g^2/\pi\rightarrow 1/\pi$.

In Fig.~\ref{fig:SPEC1}  
we show the behavior of the physical mass gap at rest, as a function of
the improved current mass $m/g$. The red-dashed line (left) is a fit to  
\bea
\label{FIT1}
\frac E{E_0}=0.99+1.76 \frac{m}{E_0}
\ee
and the green-dashed line  in the inset (left), is  a fit to
\bea
\label{FIT2}
\frac E{E_0}=\frac{0.33+1.99\, m}{E_0}.
\eea
In the inset, a clear crossing from strong to weak coupling at about $m/g\sim 1/3$ is observed, which is consistent with expectations $m/g\approx 1/\pi$ (critical point). The lightest $\eta'$ state in the spectrum is gapped by $m_s/E_0\sim 1$ for zero current mass in the strong coupling regime, and asymptotes the nonrelativistic limit of $2m/E_0\sim 2m/m_s$ in the  weak coupling regime. 

Here we emphasize  that  the spectrum of $:\mathbb H:$ following by exact diagonalization exhibits numerous spurious states, which are eliminated by imposing Gauss's law, i.e zero electric charge. More specifically, the charge operator is given by 
\begin{equation}
    Q=\sum_{n=1}^N\frac{Z_n}{2}. 
\end{equation}
and the physical eigenstates  are selected by the condition $\bra{E_n}Q\ket{E_n}=0$. Only when this condition is enforced, we see the physical cross-over reported in Fig.~\ref{fig:SPEC1}.

In Fig.~\ref{fig:boosted}, we show the
relativistic dispersion relation of boosted state obtained by exact diagonalization (dotted-blue) 
and the continuum relativistic result $p^1(v)$ (dotted-green), versus the velocity $v$ for $N=24$.
The symmetric relative error is shown in the inset in dotted-blue, where the red line indicates that it exceeds $10\%$ (at around $v\gtrsim0.83$) and the dashed-red line that it exceeds $20\%$ (at around $v\gtrsim0.91$). Moreover, increasing the mass means that the symmetric error increases faster with $v$. The orthonormality of the ground state with the boosted $\eta'$ state is only enforced for low velocities as illustrated in Fig.~\ref{fig:boostederr} (blue circles).

\section{Partonic functions from exact diagonalization}
\label{SEC7}
In this section, we will perform a numerical analysis of the quasiparton distributions in QED2 obtained by exact diagonalization as a testing ground for a future quantum simulations. In particular, we will carry out the analysis for  both the boosted form and the 'time' evolved form, to check both methods for consistency. For accuracy, the numerical results will be
checked against the exact partonic distributions in the 2-parton Fock space approximation, derived earlier.

\begin{figure*}
    \centering
    \includegraphics[width=0.45\linewidth]{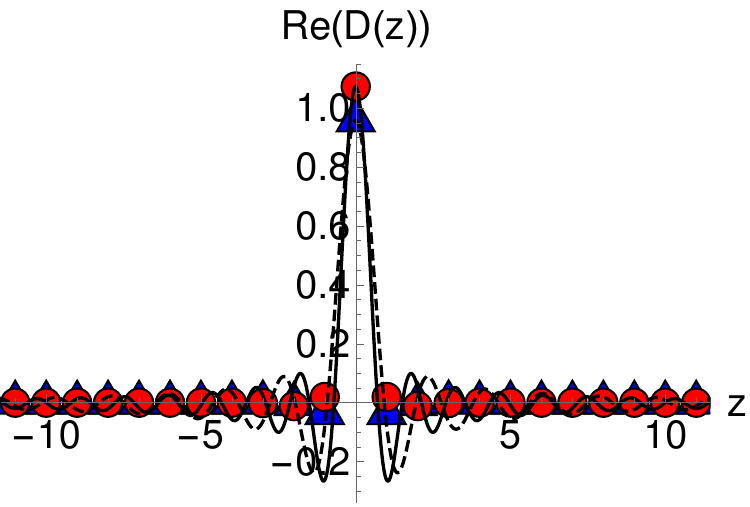}
  \includegraphics[width=0.45\linewidth]{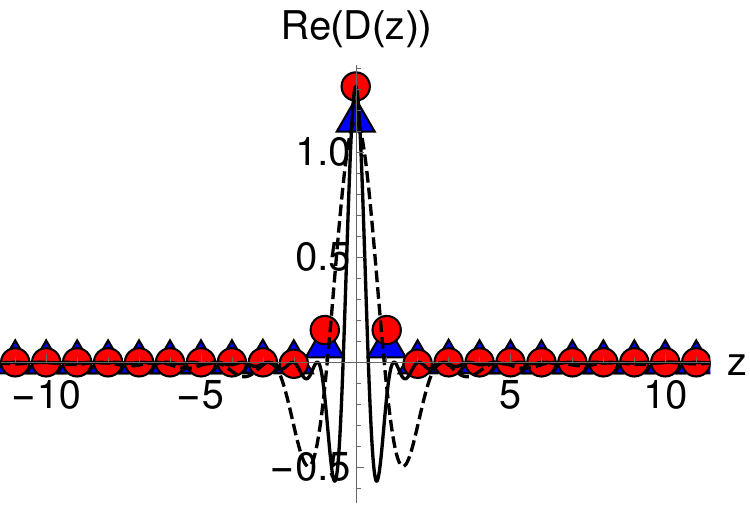}
  \includegraphics[width=0.45\linewidth]{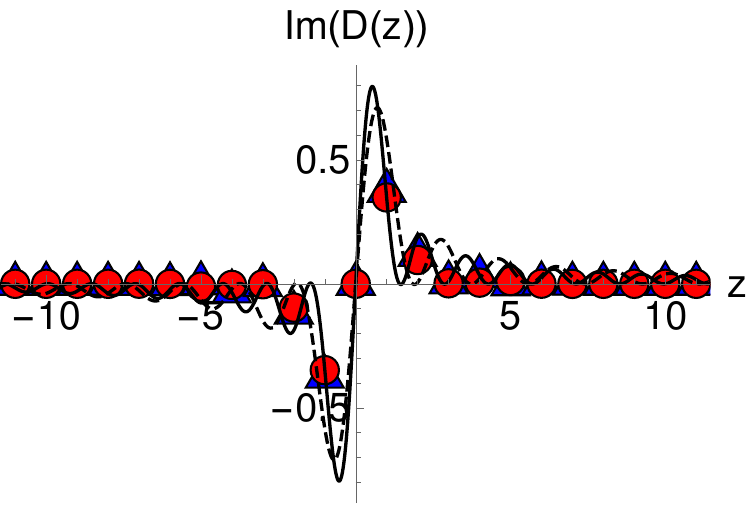}
  \includegraphics[width=0.45\linewidth]{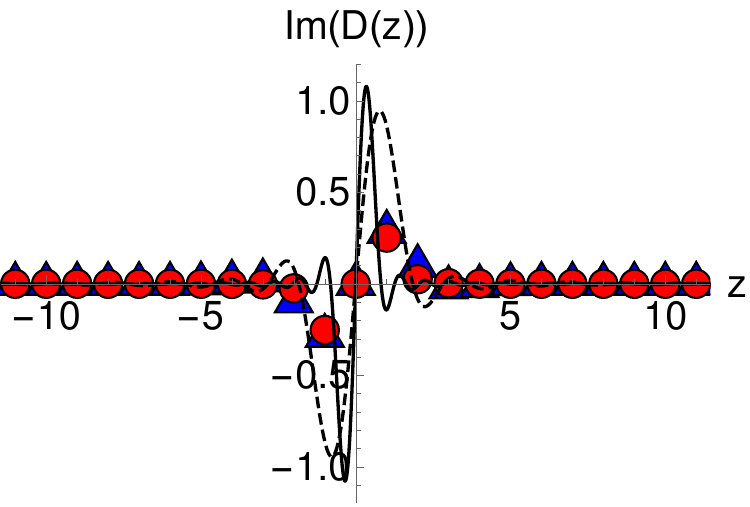}
    \caption{{\bf Left panel}: Real and Imaginary part of the spatial quasi-distribution function $D(z)$ for $v=0.925$ and $m/m_s=0.1$ (red disks)
    and the improved mass $m_\text{lat}$ (blue triangles),
    both in strong coupling.
    The exact
    result following from \eqref{TH1X} by an inverse Fourier transform is shown for $m/m_s=0.1$ (black-solid line) and the improved mass $m_\text{lat}$ \eqref{eq:mlat}
    (black-dashed line). We normalized the black curves to the peak of the real part, and fixed $N=24$, with $a=g=1.$
    {\bf Right panel}: Same as the left panel 
    but in the weak coupling regime for the improved mass  $m/m_s=0.806$.}
    \label{BOOSTED-DZ}
\end{figure*}

\subsection{Boosted form}
Using the symmetric forms (\ref{QPDA}-\ref{QPDF}), the mapped partonic distributions are
\begin{widetext}
\bea
\label{QPDA4}
\varphi_\eta(\zeta,v)\rightarrow && 
\frac {a}{4\pi\, if}
\sum_{z}\,e^{-iz\zeta P(v)}
\langle 0|
(\psi_e^\dagger(-z) +\psi_o^\dagger(-z))
(\psi_e(+z) +\psi_o(+z))\,
e^{i\chi(v) \mathbb K}\,|\eta(0 )\rangle\nonumber\\=&& 
\frac {1}{2\pi if}
\sum_{n={\rm even}}\,e^{-in\zeta aP(v)}
\langle 0|
(\varphi_{-n}^\dagger +\varphi_{-n+1}^\dagger)
(\varphi_{+n}+\varphi_{+n+1})\,
e^{i\chi(v) \mathbb K}\,|\eta(0 )\rangle\nonumber\\
\equiv&&\frac {1}{2\pi}
\sum_{n={\rm even}}\,e^{-in\zeta aP(v)}\,\varphi(na)
\eea
and 
\bea
\label{QPDF4}
q_\eta(\zeta,v)\rightarrow && 
\frac {a}{4\pi }
\sum_{z}\,e^{-iz\zeta P(v)}
\langle \eta(0)|
e^{-i\chi(v) \mathbb K}\,
(\psi_e^\dagger(-z) +\psi_o^\dagger(-z))
(\psi_e(+z) +\psi_o(+z))
\,e^{i\chi(v) \mathbb K}\,|\eta(0 )\rangle\nonumber\\=&& 
\frac {1}{2\pi}
\sum_{n={\rm even}}\,e^{-in\zeta aP(v)}
\langle \eta(0)|
e^{-i\chi(v) \mathbb K}\,
(\varphi_{-n}^\dagger +\varphi_{-n+1}^\dagger)
(\varphi_{+n}+\varphi_{+n+1})
\,e^{i\chi(v) \mathbb K}\,|\eta(0 )\rangle
\nonumber\\\equiv&& 
\frac {1}{2\pi}
\sum_{n={\rm even}}\,e^{-in\zeta aP(v)}
\,D(na).\label{eq:QPDF}\eea
\end{widetext}
with $\zeta=2x-1$. 
The rest state $|\eta(0)\rangle$ is defined as an eigenstate of $:\mathbb H:$ of lowest mass
modulo an arbitrary phase. This phase drops out in case of the distribution function, but does not in the distribution amplitude, so we set it to 1 in the latter to enforce real positivity.

Note that the spatial gauge links in (\ref{QPDA4}-\ref{eq:QPDF}) are reabsorbed into the site fermions, through the time-independent residual gauge transformation
used in mapping the Hamiltonian to its  spin form (\ref{eq:Ham}).
In 1+1 dimensions with open boundary conditions, both the electric flux and the vector potential can be eliminated, by explicitly solving Gauss' law and making use of the residual time-independent gauge transformation (modulo the flux at the left end point which we have set to 0).

\begin{figure*}
    \centering
    \includegraphics[width=0.45\linewidth]{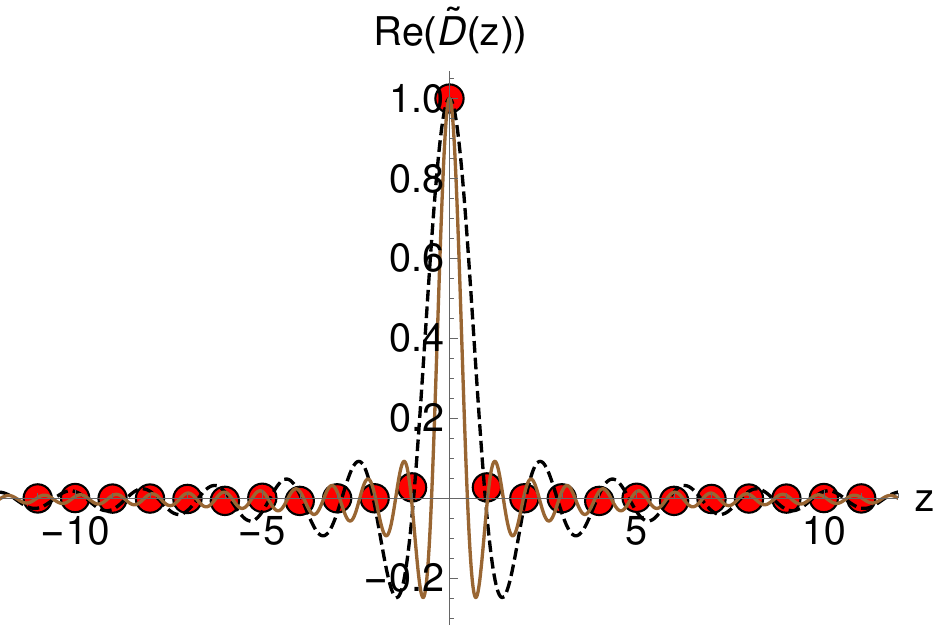}
  \includegraphics[width=0.45\linewidth]{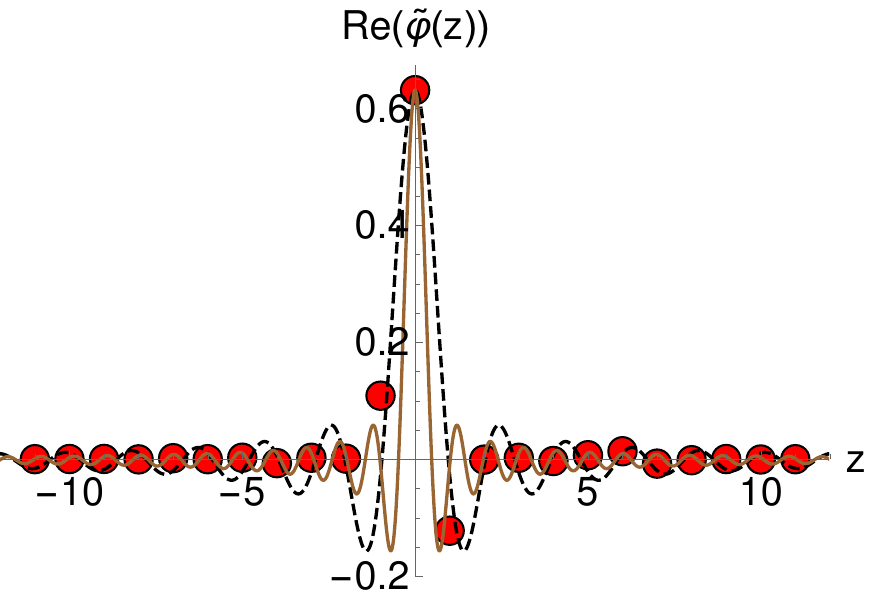}
  \includegraphics[width=0.45\linewidth]{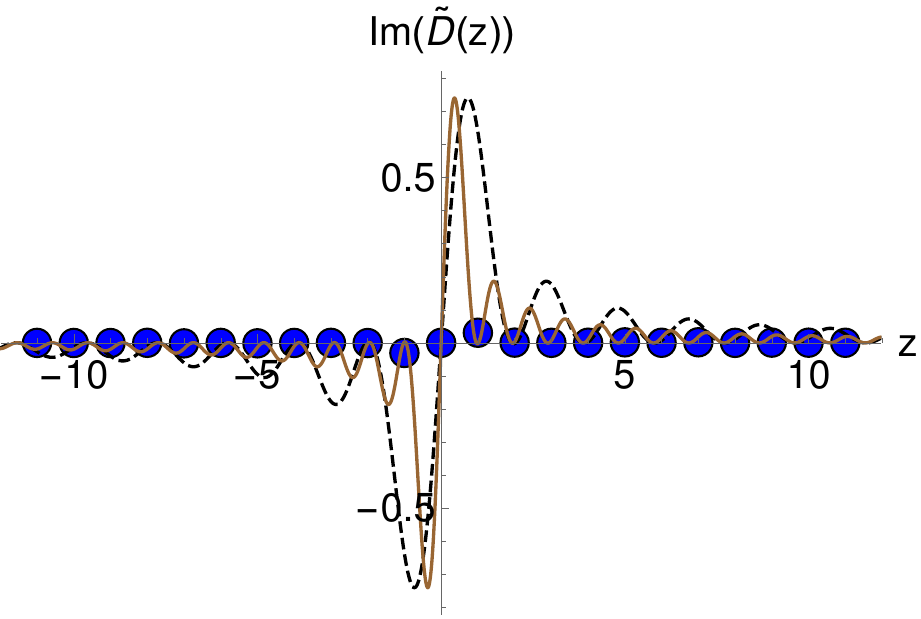}
  \includegraphics[width=0.45\linewidth]{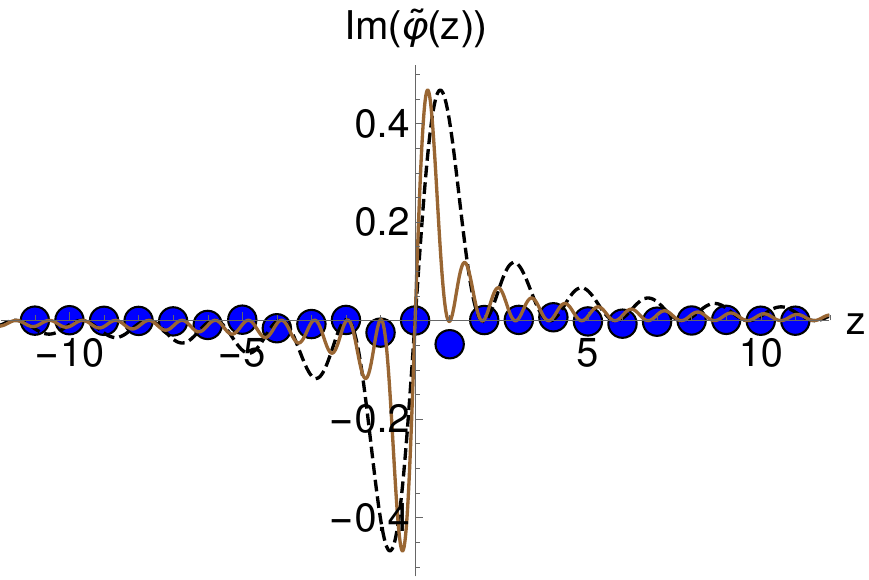}
    \caption{Real (red) and Imaginary (blue) part of $\tilde D(z)$ and $\varphi(z)$ for $v=1$ and $m/m_s=0.1$ using the improved mass,  and the time-evolved form. The black dashed line is the inverse Fourier transform (normalized to the peak of the real part) of the solution to \eqref{TH1X}. The brown line is the same function rescaled by 1/2 in $z$. In the right panel, we used the arbitrary phase between $\langle 0|$ and $|\eta\rangle$ to impose that the imaginary part vanishes for $z=0$. We fixed $N=24, a=1$ and the results are in units of $g=1$.}
    \label{TIME-DZ}
\end{figure*}

In Fig.~\ref{BOOSTED-DZ} (left-panel) we show the numerical results for the real and Imaginary  part of the spatial quasi-distribution function $D(z)$ obtained from exact diagonalization, which enters the quasiparton distribution function in \eqref{eq:QPDF}. We use the lattice configuration illustrated in Fig.~\ref{fig:lattice}. The velocity is fixed to $v=0.925$ with the current mass $m/m_s=0.1$ (red disks) and with $m_\text{lat}$ from \eqref{eq:mlat} (blue triangles), both in the weak coupling regime. The numerical results hover around zero. The inverse Fourier transformation of the exact result following from \eqref{TH1X} is shown for $m/m_s=0.1$ (black-solid line) and for $m_\text{lat}$ from \eqref{eq:mlat} (black-dashed line). 
The solid curve is normalized to the peak value of the real part. We fixed $N=24$ with $a=g=1$.

In Fig.~\ref{BOOSTED-DZ} (right-panel), we
show also the real and imaginary part of the spatial distribution of $D(z)$ for $m/m_s=0.806$ (black-dashed line) in the weak coupling regime for $N=24$ with $a=g=1$. We again fixed the black lines to the peak of the real part at $z=0$.

Finally, we note that the limited number of qubits does not allow for an approximate fit of the parton distribution function in QED2, to the inverse Fourier transform of the 2-parameter ansatz $x^\alpha(1-x)^\beta$. This procedure is commonly used in the reconstruction of the parton distributions from the spatial distributions, in current lattice simulations~\cite{Izubuchi:2019lyk}.

\begin{widetext}
\subsection{\bf 'Time' evolved  form}
Using the symmetric forms (\ref{QPDA3}-\ref{QPDF3}) with $\Gamma_v\rightarrow \gamma^+$, the mapped partonic distributions are
\bea
\label{QPDA4}
\varphi_\eta(x,v)\rightarrow && 
\frac {a}{4\pi\, i f}
\sum_{z}\,e^{-iz(\zeta-1)m_\eta v}
\langle 0|
(\psi_e^\dagger(-z) +\psi_o^\dagger(-z))\,e^{-i2vz\mathbb H}\,
(\psi_e(+z) +\psi_o(+z))\,
|\eta(0 )\rangle\nonumber\\=&& 
\frac {1}{2\pi\, if }
\sum_{n={\rm even}}\,e^{-in(\zeta-1)am_\eta v}
\langle 0|
(\varphi_{-n}^\dagger +\varphi_{-n+1}^\dagger)\,e^{-i2vn\mathbb H}\,
(\varphi_{+n}+\varphi_{+n+1})\,
|\eta(0 )\rangle\nonumber\\\equiv&& 
\frac {1}{2\pi}
\sum_{n={\rm even}}\,e^{-in\zeta am_\eta v} \tilde\varphi(na).
\eea
and 
\bea
\label{QPDF4}
q_\eta(x,v)\rightarrow && 
\frac {a}{4\pi }
\sum_{z}\,e^{-iz\zeta m_\eta v}
\langle \eta(0)|\,
(\psi_e^\dagger(-z) +\psi_o^\dagger(-z))\,
e^{-i2vz\mathbb H}\,(\psi_e(+z) +\psi_o(+z))
\,|\eta(0 )\rangle\nonumber\\=&& 
\frac {1}{2\pi}
\sum_{n={\rm even}}\,e^{-in\zeta am_\eta v}
\langle \eta(0)|\,
(\varphi_{-n}^\dagger +\varphi_{-n+1}^\dagger)
\,e^{-i2vn\mathbb H}\,
(\varphi_{+n}+\varphi_{+n+1})
\,|\eta(0 )\rangle\nonumber\\\equiv && 
\frac {1}{2\pi}
\sum_{n={\rm even}}\,e^{-in\zeta am_\eta v}
\tilde D(na).
\eea
\end{widetext}

In Fig.~\ref{TIME-DZ} (left) we show the numerical results for the 
spatial parton distributions by exact diagonalization using the "time" evolved form  (\ref{QPDA4}). We fixed $v=1$ and $m/m_s=0.1$ using the improved lattice mass \eqref{eq:mlat}, and $N=24$. The real part of the spatial distribution function $\rm Re(\tilde D(z))$ and
amplitude $\rm Re(\tilde \varphi(z))$ (red marks), are compared  to the 
exact inverse Fourier transforms before  scaling (dashed lines), and after scaling by $\frac 12$ in z (brown-solid line). In Fig.~\ref{TIME-DZ} (right) the imaginary  part of the spatial distribution function $\rm Im(\tilde D(z))$ and
amplitude $\rm Im(\tilde \varphi(z))$ (blue marks), are compared  to the inverse Fourier transforms before scaling  (dashed lines), and after scaling by $\frac 12$ in z (brown-solid line). 

The numerical results for the "time" evolved distributions, compare poorly to the exact results, in
contrast to the boosted results presented earlier. The reason maybe traced
to the use of the Hamiltonian in the ``time'' evolved form which is seen to
scale extensively with $N$, a major source of oscillations and therefore
uncertainty in the spatial distributions. This is not the case for the boost operator.

\section{Conclusions}	
\label{CONCLUSION}
We have solved massive two-dimensional QED on a discrete lattice of qubits by exact diagonalization, to analyze the 
quasiparton distributions  of the lowest meson state, both at strong and weak coupling. The digitization is performed using the standard Jordan-Wigner map of the Kogut-Susskind fermions, in a spatial lattice  with open boundary conditions. Our largest lattice consisted of $N=24$ qubits following the mapping onto the spin system.

The boosted meson state is an eigenstate of both the Hamiltonian and momentum operators in the continuum with the proper relativistic kinematics independent of the rapidity. This is not the case for the numerically obtained first excited state of our system where increasingly large deviations from  relativistic kinematics are manifest in the luminal limit.

The numerical results for the spatial quasi-distribution functions and amplitudes obtained from exact diagonalization are relatively sparse for the largest number of qubits. This is the case for both the boosted and ``time'' evolved quasi-distributions. The latter are exactly the light-cone distributions in the luminal case.
When compared to the exact inverse Fourier transformed spatial quasi-distribution functions and amplitudes, the  results show overall consistency but are far from being accurate. This is expected due to the fairly large lattice spacing.

The numerical results for the "time" evolved spatial parton distribution function and amplitude, are much less accurate than those obtained using the boosted form. This maybe traced to the large argument of the exponential in the "time" evolved form, where the Hamiltonian is seen to scale with the number of qubits, a major source of oscillations and
therefore uncertainty. This is not the case of the boost operator, which 
is less noisy since the lattice coordinate is in the weighted sum.

This exploratory analysis shows that the deviation from relativistic kinematics
of the boosted hadronic states are visible already at moderately small rapidities
and affect their orthonormality (due to finite size effects). More importantly, the oscillatory character of the 
spatial quasi-distribution functions, make them very hard to capture on a finite lattice.
To improve on this, would require much finer lattices with many more qubits than we
have used. 
Therefore, the results are encouraging and finer lattices, perhaps achieved through the use of tensor networks \cite{Florio:2023mzk} or a computation on a quantum computer, could yield good agreement with the exact results. Our approach readily extends to the analysis of  quasi-generalized parton distributions
in massive QED2. We plan to address some of these issues in upcoming work.

\begin{acknowledgments}\noindent
We thank Yahui Chai, Adrien Florio and  David Frenklakh for useful discussions.
This work is supported by the Office of Science, U.S. Department of Energy under Contract No. DE-FG-88ER40388 (SG, IZ). This research is also supported in part within the framework of the Quark-Gluon Tomography (QGT) Topical Collaboration, under contract no. DE-SC0023646. The work of SG and KI was supported by the U.S. Department of Energy, Office of Science, National Quantum Information Science Research Centers, Co-design Center for Quantum Advantage (C2QA) under Contract No.DE-SC0012704. SG was in part supported by a Feodor Lynen Research fellowship of the Alexander von Humboldt foundation.

\end{acknowledgments}

\appendix

\section{LF wavefunctions in QED2}
\label{APPNUM}
The  numerical solutions to the eigen-integral  (\ref{TH1X}) were developed 
in~\cite{Mo:1992sv}, by expanding  in a complete light front basis, which we
summarize here for completeness. More specifically, we seek
\bea
\varphi(\zeta )=\sum_n\,c_n\,f_n(\zeta ),
\eea
as an expansion using  orthonormal  Jacobi polynomials
\bea
\label{BASIS}
f_n(\zeta )=C_n\bigg(\frac {1-\zeta^2}4\bigg)^\beta\,P_n^{(2\beta,2\beta)}(\zeta)
\eea
with  the normalization
$$
C_n=\bigg(\frac{n!}2 (1+2n+4\beta)
\frac{\Gamma(1+n+4\beta)}
{\Gamma^2(1+n+2\beta)}\bigg)^{\frac 12}.
$$
The first Jacobi polynomial is constant $P_0(\zeta)=1$, which leads the solution $f_0(\zeta)$ of the form (\ref{SOL0}). This suggests that only few terms in the expansion  are needed.  
The variational parameter $\beta$ is optimally fixed by minimizing the energy. Here, we  will follow~\cite{Mo:1992sv}, and choose 
$\beta=\sqrt{3/\pi}\, (m/m_s)$ for sufficiently strong coupling.

In the basis set (\ref{BASIS}),
(\ref{TH1X}) takes a matrix form
\bea
\label{MAT1}
(\mathbb A_{mn}+\mathbb B_{mn}+\mathbb  C_{mn})\,c_n=\mu^2\,c_m,
\eea
\begin{align*}
\mathbb A_{mn}&= \frac 12\,
\int_{-1}^1d\zeta\,d\zeta' f_m(\zeta) f_n(\zeta'),\\
\mathbb B_{mn}&= 4\alpha\int_{-1}^1d\zeta\,\frac{f_m(\zeta)f_n(\zeta)}{(1-\zeta^2)},\\
\mathbb C_{mn}&=-2{\rm PP}\int_{-1}^1d\zeta\,d\zeta'\,
\frac{f_m(\zeta)(f_n(\zeta')-f_n(\zeta))}{(\zeta-\zeta')^2}.
\end{align*}
Note that the entries $\mathbb A_{mn}$ in (\ref{MAT1}) is what distinguish 
the massive Schwinger equation from the 't Hooft equation. It is due to the U(1) anomaly in QED2.

\section{x-representation}
\label{APPX}
The solutions in the x-representation with
$$x=\frac {1+\zeta}2\qquad \bar x=1-x=\frac{1-\zeta}2$$
valued in $0\leq x, \bar x\leq 1$, follow from 
\begin{widetext}
\bea
\label{TH1X}
\mu^2\varphi(x)=\int_0^1dy\,\varphi(y)+\frac {\alpha}{x\bar x}\varphi(x)-\,{\rm PP}\int_0^1dy\,\frac{\varphi(y)-\varphi(x)}{(x-y)^2}.
\eea
\end{widetext}
In this representation, the basis expansion is
\bea
\label{BASISEXP}
\varphi(x)=\sum_n \bar c_n \bar f_n(x),
\eea
with 
\bea
\label{FBARN}
\bar f_n(x)=\bar C_n(x\bar x)^\beta P_n^{(2\beta, 2\beta)}(x-\bar x).
\eea
Moreover, the normalization is given by
$$
\bar C_n=\bigg(n! (1+2n+4\beta)
\frac{\Gamma(1+n+4\beta)}
{\Gamma^2(1+n+2\beta)}\bigg)^{\frac 12}.
$$
The matrix form of (\ref{TH1X}) is
\bea
\label{MAT1X}
(\bar{\mathbb A}_{mn}+\bar{\mathbb B}_{mn}+\bar{\mathbb  C}_{mn})\,\bar c_n=\mu^2\,\bar c_m,
\eea
with the matrix entries
\begin{align}
\label{MATENT}
\bar{\mathbb A}_{mn}&= 
\int_{0}^1\,dxdy\,\bar f_m(x)\bar f_n(y),\\
\bar{\mathbb B}_{mn}&= \alpha\int_{0}^1\,dx\,\frac{\bar f_m(x)\bar f_n(x)}{x\bar x},\\
\bar{\mathbb C}_{mn}&= 
-{\rm PP}\int_{0}^1\,dxdy\,
\frac{\bar f_m(x)(\bar f_n(y)-\bar f_n(x))}{(x-y)^2}.
\end{align}

To perform the integrals analytically, we note the identity~\cite{Bardeen:1979xx}
\bea
\label{IDENT}
&&{\rm PP}\int_{0}^1\,dxdy\,
\frac{(x\bar x)^\alpha(y\bar y)^\beta}{(x-y)^2}=\nonumber\\
&&-\frac{2\pi}{2^{2(\alpha+\beta)}}
\frac{\alpha\beta\Gamma(\alpha)\Gamma(\beta)}
{(\alpha+\beta)\Gamma(\alpha+\frac 12)\Gamma(\beta+\frac 12)}
\eea
We also note that since the distribution amplitudes
$$\varphi(-\zeta)=\pm \varphi(\zeta)$$
have fixed 'parity',  the contributions
to (\ref{BASISEXP}) stem from either even or odd 
 Jacobi Polynomials. For the even polynomials
\bea
\label{POLY2}
P_{2n}^{2\beta, 2\beta}(x-\bar x)=
\sum_{m=0}^{n}\sum_{k=0}^m 
\begin{pmatrix}
    m\\
    k
\end{pmatrix}\,a_m\,(-4x\bar x)^k,
\eea
 is a polynomial in $x\bar x$, while for
 the odd polynomials behave as
\bea
\label{POLY3}
P_{2n+1}^{2\beta, 2\beta}\!(x-\!\bar x)\!=\!\!\!
\sum_{m=0}^{n}\sum_{k=0}^m \!\!
\begin{pmatrix}
    m\\
    k
\end{pmatrix}\!a_m\,(2x\!-\!1)(-4x\bar x)^k\!\!.\nonumber
\eea
 These observations, combined with (\ref{IDENT}) allow us to evaluate all matrix entries in (\ref{MATENT}) in closed analytical form. Note that in the x-representation, the parton distribution amplitude is
 \bea
 q_\eta(x)=|\varphi(x)|^2.
 \eea

\bibliography{main}
\end{document}